\documentclass[twocolumn,epsfig,showpacs,floatfix]{revtex4}
\usepackage{epsfig,colordvi}

\usepackage{epsfig}
\usepackage{graphicx}
\usepackage{graphics}
 \def\be{\begin{equation}}
 \def\ee{\end{equation}}
 \def\bea{\begin{eqnarray}}
 \def\eea{\end{eqnarray}}
 \def\bean{\begin{eqnarray*}}
 \def\eean{\end{eqnarray*}}

\begin{document}
\title{Is the centrality dependence of the elliptic flow $v_2$ and
of the average $<p_T>$ more than a Core-Corona Effect?}
\author{J. Aichelin and K. Werner}
\affiliation{SUBATECH, Laboratoire de Physique Subatomique et des
Technologies Associ\'ees, \\
Universit\'e de Nantes - IN2P3/CNRS - Ecole des Mines de Nantes \\
4 rue Alfred Kastler, F-44072 Nantes, Cedex 03, France\\}
\begin{abstract}Recently we have shown that the centrality dependence of the
multiplicity of different hadron species observed in RHIC and SPS experiments can be well understood in a simple model,
dubbed core-corona model. There it is assumed that those incoming nucleons which scatter only once produce
hadrons as in pp collisions whereas those which scatter more often form an equilibrated
source which decays according to phase space. In this article we show that also kinematical
variables like $v_2/\epsilon_{part} (N_{part})$ as well as $v_2^i/\epsilon_{part} (N_{part})$ and $<p_T^i (N_{part})>$
of identified particles are well described in this model. The correlation of $<p_T^i>$ between peripheral heavy ion collisions
and pp collisions for different hadrons, reproduced in this model, questions whether hydrodynamical calculations
are the proper tool to describe non-central heavy ion collision. The model explains as well the centrality
dependence of $v_2/\epsilon_{part}$ of charged particles, considered up to now as an observable which allows to determine the viscosity of the quark gluon plasma. The observed dependence of $v_2^i/\epsilon_{part}(N_{part})$ on the particle species
is a simple consequence of the different ratios of core to corona particles.

\end{abstract}
\date{\today} \maketitle
\begin{figure}[ht]
\begin{center}
\hspace*{-0.5cm}
\includegraphics[width=9.5cm]{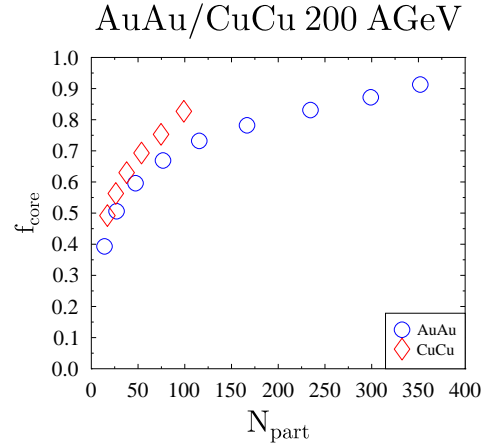}
\end{center}
\caption{$f_{core}$ the fraction of core nucleons as a function of the participant number $N_{part}$ for AuAu and CuCu collisions at $\sqrt{s}=200 GeV$.} \label{ex99}
\end{figure}

Simulations of heavy ion collisions with advanced event generators like
EPOS \cite{klaus}, which reproduce a multitude of experimental observables, have revealed that nucleons at the surface of the reaction zone (called corona particles) have only few collisions and do not come to statistical equilibrium with the more central (core) particles which form an equilibrated system. To study the consequences of this observation we have developed a simple model \cite{Aichelin:2008mi} by defining corona particles as those nucleons which have only one initial collision whereas the others are considered as core
particles. $f_{core}$ is the fraction of core nucleons which depends on the centrality, the system size and (weakly) on the beam energy. In this simple model we could show that, independent of the system size, the centrality dependence of the multiplicity of all hadrons from SPS to RHIC energies can quantitatively be described  by:
\begin{eqnarray}
 & &M^i(N_{\rm part})\label{eq1}\\
 & & = N_{\rm part}\,\big[f_{core}\cdot M^i_{\rm core}+(1-f_{core})\cdot M^i_{\rm corona}\big]
\,\nonumber\end{eqnarray}
where $f_{core}$, shown in Fig. \ref{ex99}, has been calculated in a Glauber model and $i$ refers to the hadron species.
For our calculation we fix $M_{\rm core}^i$ by applying eq. \ref{eq1} to the
most central AuAu or PbPb data point. $M_{\rm corona}^i$ is
given as half of the multiplicity measured in pp collisions. Once these
parameters are fixed, the centrality dependence of $M^i$ is
determined by eq. \ref{eq1}. Especially the centrality dependence of
the lighter CuCu system follows then without any further input. Certainly
this is a very simple model with no free parameter (besides the Glauber calculation of
$f_{core}$) but the present experimental error bars of the quantities which we analyze give not
sufficient information to refine the model.

One may ask whether the core - corona model can also describe the centrality dependence of other observables, like $<p_T>(N_{part})$ or $v_2/\epsilon_{part} (N_{part})$ of charged particles or identified hadrons. Especially the centrality dependence of
$v_2/\epsilon_{part} (N_{part})$ has recently created a lot of theoretical activities.
Initially the azimuthal distribution $d\sigma/d\phi \propto (1+2 v_1\cdot \cos(\phi) + 2 v_2\cdot \cos(2\phi))$ of light quarks and gluons is isotropic and the anisotropy develops during the expansion as an image of the initial eccentricity $\epsilon_{part}$ in coordinate space. This initial eccentricity has been calculated in a Glauber model or in a Color Glass Condensate approach. Taking this eccentricity as input, calculations based
on ideal hydrodynamics reproduce the $v_2/\epsilon_{part}$ observed in central AuAu
collisions at RHIC. They predict, however, a centrality independent $v_2/\epsilon_{part}$ value and fail therefore to describe the measured centrality dependence of $v_2/\epsilon_{part}$ which decreases by a factor of two from central to peripheral reactions.

Drescher et al. \cite{Drescher:2007cd} argue that this centrality dependence is a sign that the parton cross section in the quark gluon plasma, created presumably at RHIC energies, is too small in order to justify the application of ideal hydrodynamics. It has to
be replaced by viscous hydrodynamics. They could show that the impact parameter dependence of $v_2/\epsilon_{part}$ can be related to the Knudsen number
and therefore to a finite viscosity. Later Luzum et al. \cite{Luzum:2008cw}
showed in viscous hydrodynamical calculations that different viscosities yield a different centrality dependence of $v_2/\epsilon_{part}$ and of $<p_T>$. Performing calculations with different values for the viscosity they could identify the viscosity which fits best the 200 AGeV AuAu data and propose that this value is characteristic for the quark gluon plasma. More recently, Song et al. \cite{Song:2008si} have extended the calculation to the CuCu system.
It is therefore challenging to explore the predictions of the core - corona model for dynamical observables and to compare the results to data.

The centrality dependence of $<p_T^i>$ at midrapidity predicted by the core - corona model is given by
\begin{eqnarray}
 & &<p_T^i>(N_{\rm part})\label{eq2}\\
 & & = f_{core}^i\cdot <p_T^i>_{\rm core}+(1-f_{core}^i)\cdot <p_T^i>_{\rm corona}
\, . \nonumber\end{eqnarray}
where i characterizes the particle species i. $<p_T^i>_{\rm corona}$ is the value of $<p_T^i>$ measured in pp
collisions at midrapidity whereas $<p_T^i>_{\rm core}$ is the value of $<p_T^i>$ measured in central heavy ion reactions at midrapidity. $f_{core}^i$ is the centrality dependent fraction of core particles for the particles species i, given by
\be
f_{core}^i= \frac{f_{core}\cdot M^i_{\rm core}}{
f_{core}\cdot M^i_{\rm core}+(1-f_{core})\cdot M^i_{\rm corona}}.
\label{eq3}
\ee
If the system consists of particles of one species, final state interactions change neither $<p_T>$ nor  $v_2/\epsilon_{part}$ substantially. The former is due to (transverse) energy conservation, the latter due to the small momentum transfer in final state interactions as compared to the momentum of the particles. It is also not important whether some corona particles are absorbed by the core. This increases the core fraction but simultaneously lowers $<p_T>_{core}$. Both effects compensate as long as the total transverse energy is conserved.

This observation is not anymore true if the systems contains different particles species i. As we will see, the
dependence of  $<p_T^i>$ on $N_{part}$ is quite different for different hadrons. Interactions transfer momentum between the different particles species and hence change $<p_T^i>$ in the direction toward a thermal equilibrium.  Therefore we do not expect that $<p_T^i>$ follows eq.\ref{eq2} if final state interactions between the hadrons become important.

Consequently, if eq. \ref{eq2} is in agreement with data, we can draw conclusions on the reaction scenario.

The NA49 collaboration \cite{Anticic:2009wd} has recently analyzed the transverse mass spectrum of different
hadrons in the framework of our model and has found good agreement with data.
Therefore it is meaningful to continue here this analysis towards higher energies.

\begin{figure}[ht]
\begin{center}
\vspace*{-0.5cm}
\includegraphics[width=9.5cm]{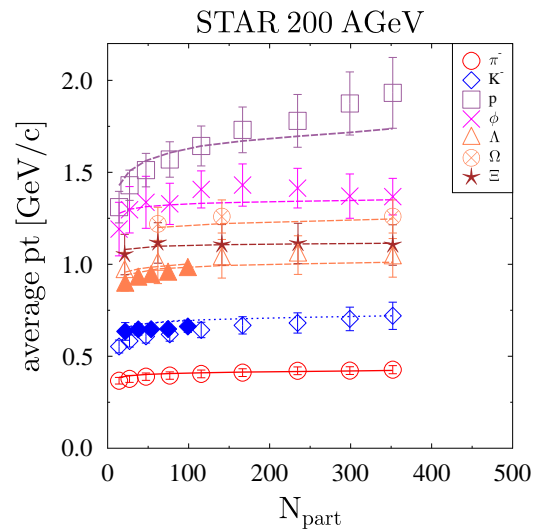}
\end{center}
\caption{$<p_T^i>$ as a function of the centrality for different particle species i.
The STAR data \cite{star:2008ez,Abelev:2008zk,star:2008fd,mag,tim} are presented by symbols, the calculations, eq.\ref{eq2}, by lines. Full (open) symbols refer to CuCu (AuAu). The $<p_T>$ values of protons ($\phi$) are multiplied by a factor of 1.75(1.4).} \label{pthad}
\end{figure}
\begin{figure}[ht]
\begin{center}
\vspace*{-0.5cm}
\includegraphics[width=9.5cm]{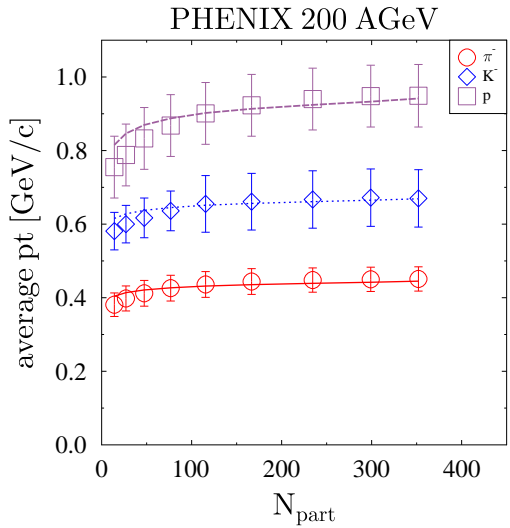}
\end{center}
\caption{$<p_T^i>$ as a function of the centrality for different particle species i.
The PHENIX data \cite{Adler:2003cb} are presented by symbols, the calculations by lines.} \label{pthadp}
\end{figure}

In  fig. \ref{pthad} we compare our predictions with published \cite{star:2008ez,star:2008fd,Abelev:2008zk} and  preliminary
\cite{mag,tim} STAR data, in fig. \ref{pthadp} with PHENIX data \cite{Adler:2003cb}.
In  fig. \ref{pthad} the theoretical and experimental values of protons
($\phi$) are multiplied by a factor of 1.75(1.4) to separate the curves. The STAR proton data show
(in contradistinction to the proton data of PHENIX and to those of all other hadrons) a strong increase of
$<p_T>$ for $N_{part} > 100$. We have therefore multiplied the experimental $<p_T>_{core}$ by 0.9.

We see that $<p_T^i(N_{part})>$, calculated with help of eq. \ref{eq2}, agrees quantitatively with the data
for the different hadrons. Also the experimentally observed different dependencies of $<p_T^i(N_{part})>$
for different particle species i is well reproduced.
We note in passing that replacing $f_{core}^i$ by $f_{core}$ in eq. \ref{eq2} changes the curve only slightly.
Its form is dominated by the difference between  $<p_T>_{core}$ and  $<p_T>_{corona}$.
Tab.\ref{tab:meanpt} summarizes the experimental $<p_T^i>_{\rm corona}$ and $<p_T^i>_{\rm core}$ which enter our calculations and which are taken from pp and central heavy ion collisions. PHENIX has measured very peripheral heavy ion collisions. The $<p_T>$ value there coincide within the error bars with the STAR pp data. We note that  $<p_T^i>_{\rm corona}/<p_T^i>_{\rm core}$ varies from .81 for pions to .62 for protons.
\begin{table}[!ht]
\begin{center}
\begin{tabular}{|c|c|c|c|c|c|c|c|c|}
\hline
&\multicolumn{4}{c|}{STAR}& \multicolumn{2}{c|}{PHENIX}\\
&\multicolumn{2}{c|}{pp}& \multicolumn{2}{c|}{$AA_{central}$}&\multicolumn{2}{c|}{$AA_{central}$}\\
\hline \textbf{Particle}    &
       $\mathbf{\langle{\bf p_{T}}\rangle}$ &
       \textbf{Err.} &$\mathbf{\langle{\bf p_{T}}\rangle}$ &
       \textbf{Err.} &$\mathbf{\langle{\bf p_{T}}\rangle}$ &
       \textbf{Err.} \tabularnewline\hline
$\pi^-$ &0.348& 	0.018& 0.429& 	0.022& 0.455 &0.032 \\
$K^-$ &	0.605& 	0.073& 0.717& 	0.050 & 0.677& 0.068 \\
$K_0$ &	0.600& 	0.030& 0.800& 	0.050 &&\\
$p$&0.686&0.041&1.104&0.11&0.949&0.085\\
${\bar p}$& 	0.683& 	0.041& 1.105& 	0.069 & 0.959 & 0.084\\
$\phi$& 	0.820& 	0.050& 0.970& 	0.097&&\\
$\Lambda$& 	0.780&	0.040& 1.050& 	0.090&&\\
${\bar\Lambda}$& 	0.760& 	0.040&&&&\\
$\Xi$& 	0.924& 	0.130& 1.110& 	0.090&&\\
${\bar \Xi}$& 	0.881& 	0.130&&&&\\
$\Omega+{\bar \Omega}$& 	1.080& 	0.300& 1.260& 	0.090&&\\  \hline
\end{tabular}
\caption{A summary of mid-rapidity $\langle{\bf p_{T}}\rangle$ in GeV/c for identified particles from the STAR \cite{star:2008ez,star:2008fd,Abelev:2008zk,mag,tim} and the PHENIX  \cite{Adler:2003cb} collaboration.}
\label{tab:meanpt}
\end{center}
\end{table}

\begin{figure}[ht]
\begin{center}
\hspace*{-0.5cm}
\includegraphics[width=9.5cm]{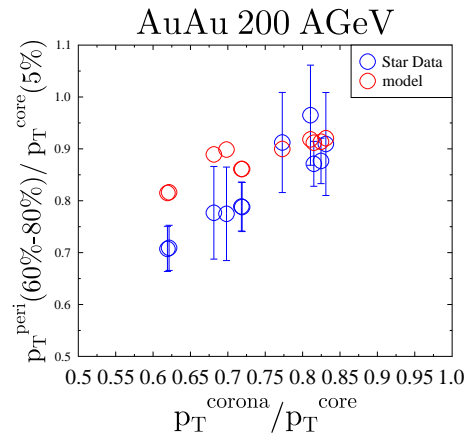}
\end{center}
\caption{The ratio of $<p_T>$ in peripheral to central reactions as compared to that
of pp to central reactions. The data are from \cite{Abelev:2006cs, star:2008fd, Abelev:2008zk,mag,tim, star:2008ez}.
Because the calculation uses
the experimental data, the error bars of model predictions are of about the same size as that of the experimental points.
}
\label{centper}
\end{figure}

By construction the core - corona model correlates the
$<p_T^i>$ values of peripheral heavy ion reactions with those observed in pp.
This correlation is shown in fig.\ref{centper}. It displays $<p_T^i>_{peri}/<p_T^i>_{core}$ as a function of $<p_T^i>_{corona}/<p_T^i>_{core}$, using eq.\ref{eq2}, for different hadron species.
The experiential data show this correlation as well, it is even stronger there.
Thus also in experiment the value of $<p_T^i>$ in peripheral reactions is small as compared to that
in central collisions for those hadrons for which $<p_T^i>$ measured in pp is small as compared to $<p_T^i>$ measured in central heavy ion collisions.
This correlation is far from being trivial. If, as often assumed, the heavy ion collision can be described by
hydrodynamical calculations we do not expect any correlation between peripheral heavy ion data and pp data.
Therefore the correlations observed in fig. \ref{centper} support the validity of the core - corona approach and questions
whether hydrodynamical calculations are the proper tool to describe non-central heavy ion reactions.

\begin{figure}[ht]
\begin{center}
\hspace*{-0.5cm}
\includegraphics[width=9.5cm]{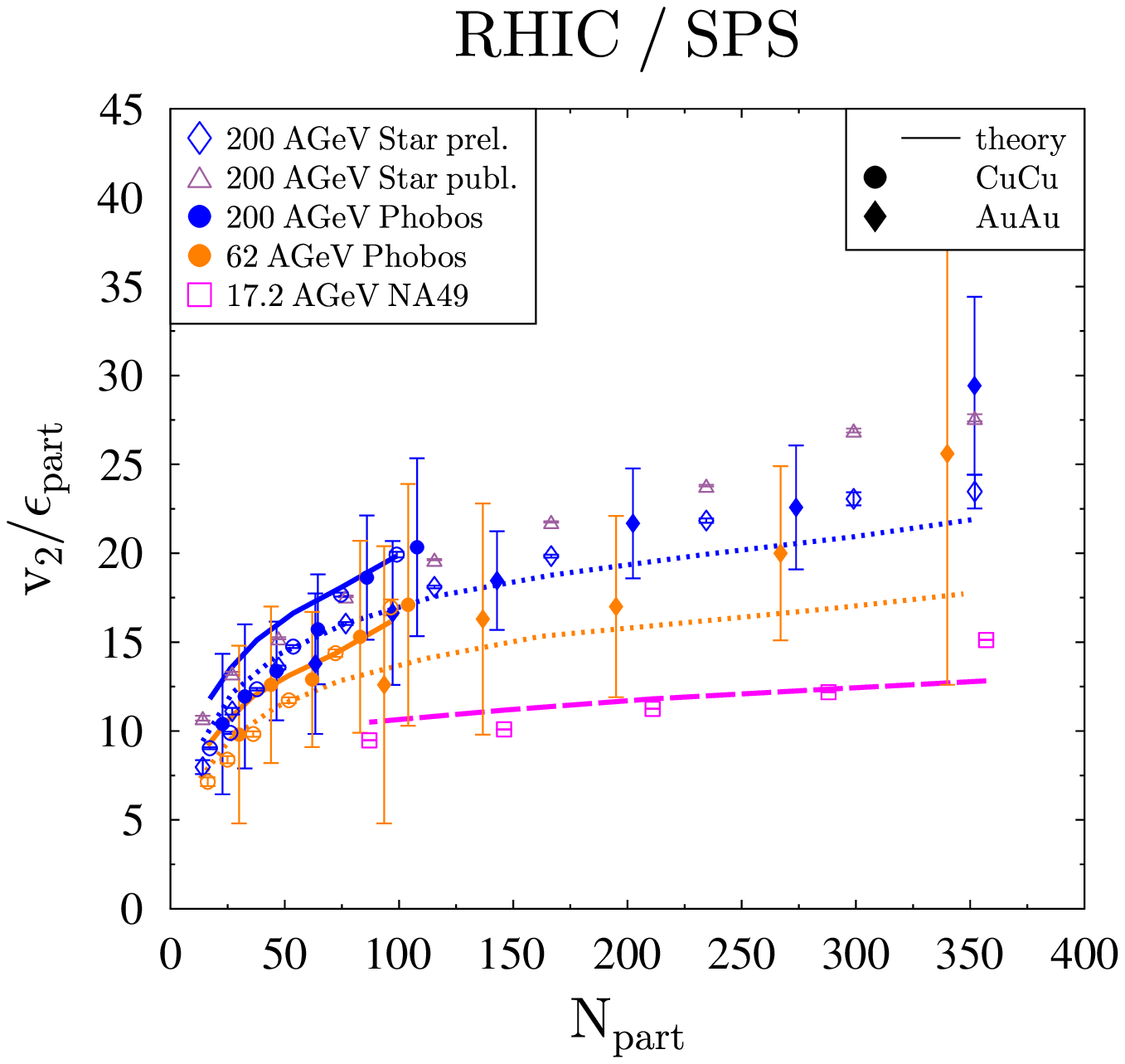}
\includegraphics[width=9.5cm]{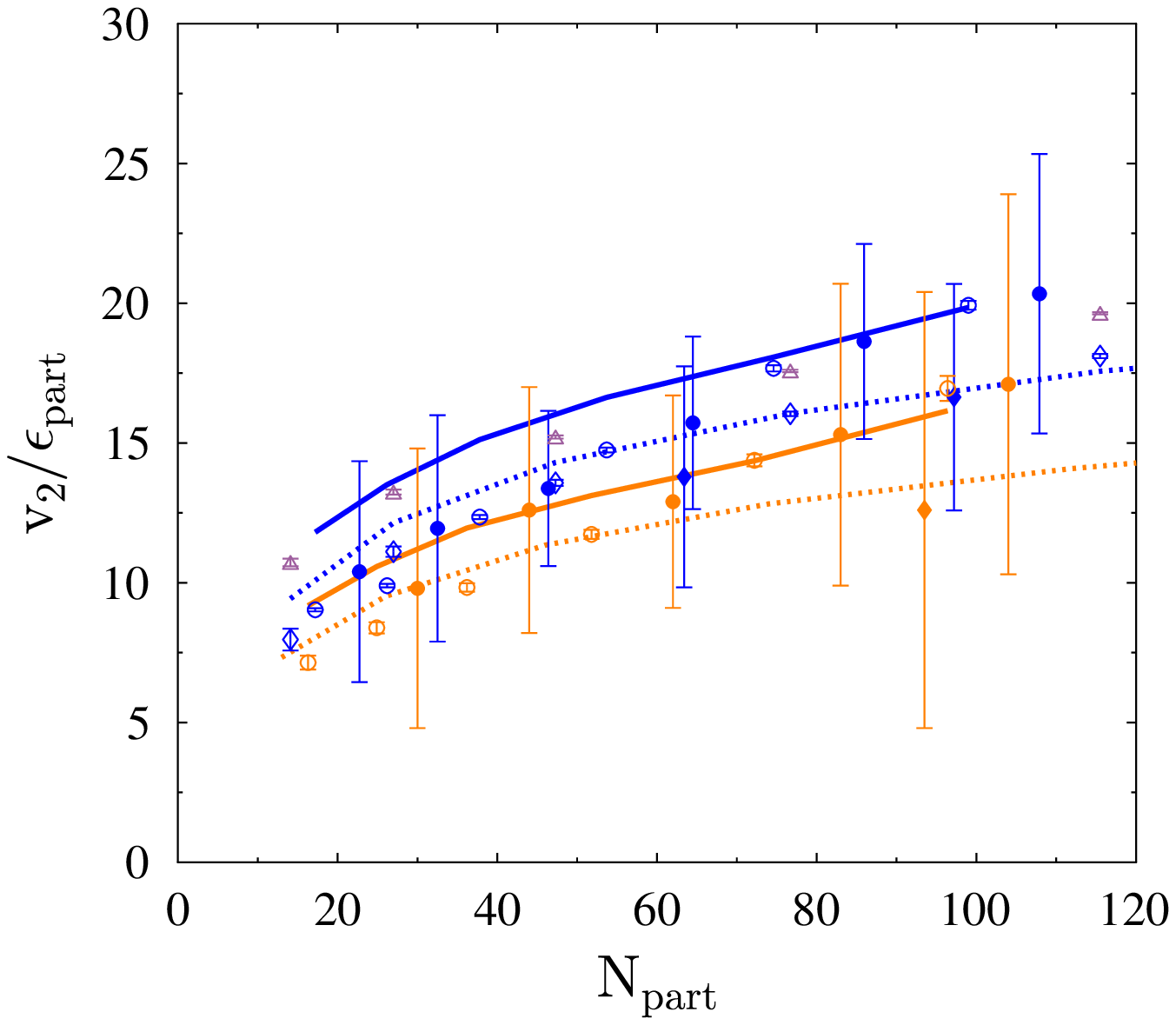}
\end{center}
\caption{Top: $N_{part}$ dependence of $v_2/\epsilon_{part}$ for charged hadrons in AuAu and CuCu at 200 AGeV and
62 AGeV as measured by PHOBOS\cite{Alver:2006wh} and  STAR \cite{star:2008ed, shi} as well as at 17.2 AGeV \cite{Alt:2003ab} as measured  by NA49 in comparison with theory. Bottom:
Zoom of the region of small $N_{part}$.}
\label{v2e}
\end{figure}

\begin{figure}
\begin{center}
\hspace*{-0.5cm}
\includegraphics[width=9.5cm]{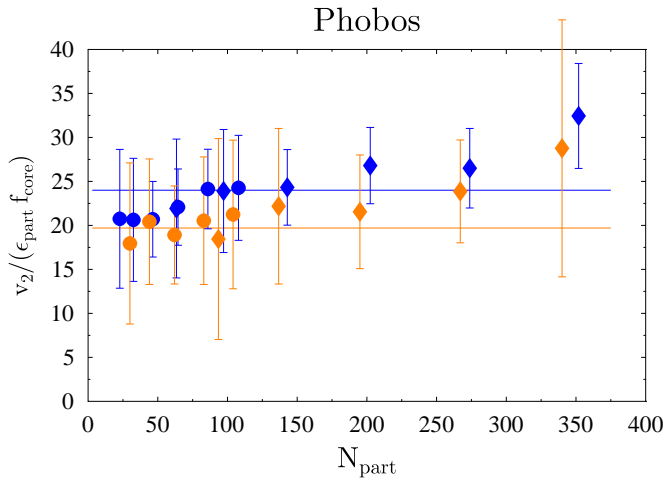}
\end{center}
\caption{$N_{part}$ dependence of $v_2/(\epsilon_{part} f_{core}) $ for charged hadrons in AuAu and CuCu at 200 AGeV/62 AGeV measured by the PHOBOS collaboration\cite{Alver:2006wh} as compared to theory. } \label{v2ef}
\end{figure}

Corona particles do not form a thermal equilibrium with the core particles but decay
like pp collisions and hence isotropically. Only the core particles feel the eccentricity of the overlap region.
Because  $M_{core}^{charged\ particles} \approx M_{corona}^{charged\ particles}$,
$v_2/\epsilon_{part}$ is expected to be $\propto f_{core}$ and therefore the
expected centrality dependence of $v_2/\epsilon_{part}$ is that of $f_{core}$.  This allows for some immediate predictions. Because $f_{core}$ is similar in central CuCu and AuAu collisions (Fig. \ref{ex99}) we expect that central collisions of AuAu and CuCu show a similar $v_2/\epsilon_{part}$ despite of the large difference of $N_{part}$. In peripheral CuCu and AuAu collision $v_2/\epsilon_{part}$ should be very similar for the same $N_{part}$. In order to compare our model with experiment we have to determine $(v_2/\epsilon_{part})^{hydro}$, i.e. $v_2/\epsilon_{part}$ for $f_{core}=1$.
We take $(v_2/\epsilon_{part})^{hydro}$ as 24\%, 19.7\%  and 12.6\% for reactions at $\sqrt{s}$ of 200
AGeV, 62 AGeV and 17.2 AGeV, respectively. In fig.\ref{v2e} we display the experimental results from PHOBOS (full symbols), from STAR (open symbols) and from NA49 (open squares) in comparison with our predictions, presented as lines. We see a quite nice agreement. As in our model, for a given number of $N_{part}$ the experimental values of $v_2/\epsilon_{part}$ for CuCu are higher than those for AuAu. Hydrodynamical calculations predict the opposite. This allows to conclude on the underlying mechanism once the error bars of the present preliminary data are further reduced.
The situation becomes much more clear when
we divide $v_2/\epsilon_{part}$
by $f_{core}$. Our model predicts a constant. Fig. \ref{v2ef} presents  the
PHOBOS data for the 4
different systems, AuAu and CuCu at 200 and 64 AGeV .
The experimental $v_2/(\epsilon_{part} f_{core}) $ is indeed independent of $N_{part}$ which means that the centrality dependence of $v_2/(\epsilon_{part}) $ agrees with that of $f_{core}$. This does not exclude, of course, that in hydrodynamical calculations a viscosity can be obtained by fitting the centrality dependence of the experimental data. In our approach we do, however,  not need such a fit parameter. It is astonishing that even for the most peripheral reactions, where the number of core particles is of the order of
10 and therefore the assumption of a hydrodynamical elliptical flow is questionable, the predictions of  core - corona model still hold.
\begin{figure}
\begin{center}
\hspace*{-0.5cm}
\includegraphics[width=9.5cm]{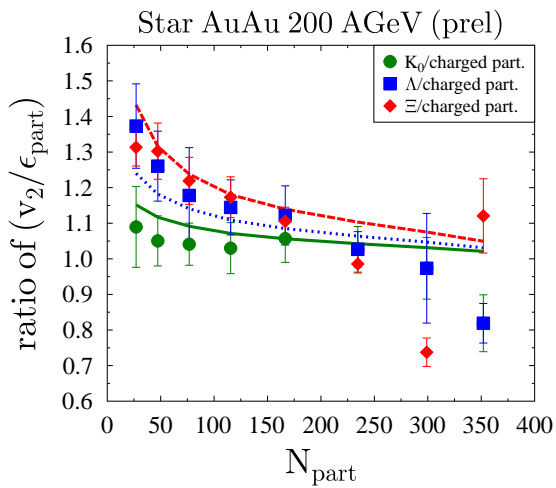}
\end{center}
\caption{Centrality dependence of $\frac{v_2^i}{\epsilon_{part}} (N_{part})/ \frac{v_2^{charged \ hadrons}}{\epsilon_{part}} (N_{part})$ for the preliminary $p_T$ integrated  $K_0, \Lambda, \Xi$ data as measured by the STAR collaboration \cite{shi} in comparison with the prediction of the core-corona model, eq.\ref{eq4}, presented as lines.}
\label{v2had}
\end{figure}
\begin{figure}
\begin{center}
\hspace*{-0.5cm}
\includegraphics[width=9.5cm, angle=270]{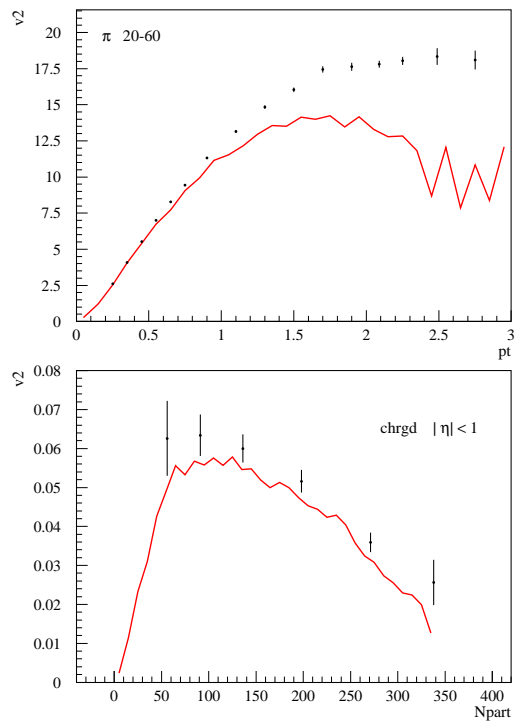}
\end{center}
\caption{EPOS calculation compared with data. The top panel shows $v_2$ as a function of $p_T$ for
$\pi^+$ measured in semi-central reactions. The bottom panel shows the $p_T$ integrated
value of $v_2$ as a function of the number of participants. Data are from ref. \cite{shi}.} \label{joerg}
\end{figure}

In our approach $v_2/\epsilon_{part}$ is a property of the plasma particles and therefore one may expect the same
centrality dependence of $v_2/\epsilon_{part}$ for all hadrons (although we cannot exclude that the hadronization itself
creates differences for different types of hadrons). This is, however, not true. The fraction of particles which come
from the core depends, according to eq. \ref{eq1}, on the particle species. For all
charged particles and $\pi$, for which $M_{\rm core}\approx M_{\rm corona}$, this dependence is weak \cite{Adler:2003cb} (and therefore we have divided $v_2$ by $f_{core}$) but for multi-strange hadrons, for which $M_{\rm core} >> M_{\rm corona}$, this is an important effect. We expect
 \begin{eqnarray}
 & &\frac{v_2^i}{\epsilon_{part}} (N_{part}) \label{eq4}\\
 & & = \frac{f_{core}\cdot M^i_{\rm core}}{f_{core}\cdot M^i_{\rm core}+(1-f_{core})\cdot M^i_{\rm corona}}\cdot \big(\frac{v_2}{\epsilon_{part}}\big)^{hydro}.
\,\nonumber\end{eqnarray}
Fig.\ref{v2had} shows the centrality dependence of $\frac{v_2^i}{\epsilon_{part}} (N_{part})/ \frac{v_2^{charged \ hadrons}}{\epsilon_{part}} (N_{part})$ for the preliminary $p_T$ integrated  $K_0, \Lambda, \Xi$ data as measured by the STAR collaboration \cite{shi}.
These data are compared with the model, eq. \ref{eq4}. Model and experiment agree within error bars and the trend of the data is well reproduced.
We note that the $p_T$ integrated value of $v_2$ does not show a scaling with the number of entrained quarks $n_q$ which has been observed in the differential spectra for $(m_T-m_0)/n_q  < 1 GeV$, with $m_T$ $(m_0)$ being the transverse (rest) mass.
The relative ratio of $\frac{v_2^i}{\epsilon_{part}}$ between baryons and mesons depend on centrality and on the particle species and is not 3/2 corresponding to the ratio of entrained constituent quarks. However, before firm conclusions can be drawn
more data have to be available. Especially data for protons for which
$f_{core}^i$ and $f_{corona}^i$ differ much less than for the multi-strange hadrons would be of help.
Such an analysis will not be easy as can be seen from fig.
\ref{joerg} which shows $v_2$  of $\pi$'s as a function of $p_T$ for semi-central reactions
(20\%-60\% centrality) (top panel) and that of charged particles as a function of $N_{part}$ (bottom panel).
We compare results of EPOS calculations with the experimental data.
The top panel shows that the  $v_2(p_T)$ spectra is qualitatively described up to $p_T \approx 1 GeV/c$, the range where almost all particles are located. Nevertheless, after integration over $p_T$, bottom panel, we see differences of $v_2$ of the order of 10\% between theory and experiment, although $v_2$ of $\pi$'s and of charged particles should be very similar. This is due to the extrapolation of the experimentally measured curve toward $p_T \to 0$, necessary to determine a $p_T$ integrated $v_2$ value.

Summarizing, the centrality dependence of all observables for identified particles, their multiplicity, their average transverse momentum and their elliptical flow, as well as the elliptical flow of charged hadrons can be quantitatively described from SPS to RHIC energies in the very simple core - corona model. It would be interesting to compare these results in detail with the predictions of (viscous) hydrodynamical calculations, the alternative approach to describe non central heavy ion collisions.

 The model predicts even the experimentally observed correlations between peripheral heavy ion and pp collisions. Such a correlation is alien to any hydrodynamical approach. Also the larger $v_2/\epsilon_{part}$ value for CuCu as compared to AuAu for the same number of $N_{part}$ questions whether the centrality dependence of $v_2/\epsilon_{part}$ can be related to the viscosity of the plasma.

The success of the core - corona model suggests the following reaction scenario: The
experimental data are a superposition of two contributions: a corona contribution which is given by pp
physics and a core contribution which agrees with the assumption that the core forms an equilibrated source.
The data are compatible with the assumption that the properties of this source are independent of centrality.
A centrality dependence of observables appears because the relative fraction of core and corona contribution varies with centrality.
Core particles are in statistical  equilibrium. Their interaction with corona particles as well as the interaction among corona particles
has to be unimportant otherwise momentum would be transferred between the different species and eq. \ref{eq2} would not be valid anymore. This suggest also a very limited number of collisional interactions between hadrons from core and corona.

Acknowledgements: We would like to thank Drs. R. Bellwied, M. Estienne, U. Heinz, J.Y. Ollitrault, S. Shi, R. Snellings,  A. Timmings for valuable discussions.

\end{document}